# Charged Impurity Scattering in Graphene

J.H. Chen[1,4*], C. Jang[1,2,4*], M. S. Fuhrer[1,2,4], and E. D. Williams[1,3,4,5], and M. Ishigami[4]

*[1]Materials Research Science and Engineering Center, [2]Center for Nanophysics and Advanced Materials, [3]Laboratory for Physical Sciences, [4]Physics Department, [5]Institute for Physical Science and Technology, University of Maryland, College Park, MD 20742 USA*

*These authors contributed equally to this work.

Since the experimental realization of graphene[1], extensive theoretical work has focused on short-range disorder[2-5], "ripples"[6,7], or charged impurities[2,3,8-13] to explain the conductivity as a function of carrier density $\sigma(n)$[1,14-18], and its minimum value $\sigma_{min}$ near twice the conductance quantum $4e^2/h$[14, 15, 19, 20].  Here we vary the density of charged impurities $n_{imp}$ on clean graphene[21] by deposition of potassium in ultra high vacuum.  At non-zero carrier density, charged impurity scattering produces the ubiquitously observed[1, 14-18] linear $\sigma(n)$ with the theoretically-predicted magnitude. The predicted asymmetry[11] for attractive vs. repulsive scattering of Dirac fermions is observed.  $\sigma_{min}$ occurs not at the carrier density which neutralizes $n_{imp}$, but rather the carrier density at which the average impurity potential is zero[10].  $\sigma_{min}$ decreases initially with $n_{imp}$, reaching a minimum near $4e^2/h$ at non-zero $n_{imp}$, indicating that $\sigma_{min}$ in present experimental samples does not probe Dirac-point physics[14, 15, 19, 20] but rather carrier density inhomogeneity due to the impurity potential[3, 9, 10].

Figure 1a shows an optical micrograph of the graphene device used in this study, and Figure 1b shows its micro-Raman spectrum; the single Lorentzian D' peak confirms that the device is single-layer graphene[22] (see Methods). To vary the density of charged impurities, the device was dosed with a controlled potassium flux in sequential 2-second intervals at a sample temperature $T$ = 20 K in ultra-high vacuum (UHV). The gate-voltage-dependent conductivity σ($V_g$) was measured *in situ* for the pristine device, and again after each doping interval. After several doping intervals, the device was annealed in UHV to 490 K to remove weakly adsorbed potassium[23], then cooled to 20 K and the doping experiment repeated; four such runs (Runs 1-4) were performed in total.

Figure 2 shows the conductivity vs. gate voltage for the pristine[21] device and at three different doping concentrations at 20K in UHV for Run 3 (see also Supplemental Fig. S2, and Fig. S3 for measurements on a second device). Several features are notable immediately in Figure 2 and comprise the major experimental observations in this work. Upon K-doping, (1) the mobility decreases, (2) the gate-voltage dependence of the conductivity σ($V_g$) becomes more linear, (3) the mobility becomes larger for holes than electrons, (4) the gate voltage of minimum conductivity $V_{g,min}$ shifts to more negative gate voltage, (5) the width of the minimum conductivity region in $V_g$ broadens, and (6) the minimum conductivity σ$_{min}$ decreases, at least initially. In addition, (7) the linear σ($V_g$) curves extrapolate to a finite value at the minimum conductivity point, as discussed below. All of these features have been predicted[2, 3, 8-10, 12, 13] for charged impurity scattering in graphene, we will discuss each in detail below. We note that a previous study on chemical doping under poorly controlled adsorption conditions[24] reported effects (4) and (5) without changes in mobility (see Supplementary Note).

We first examine the behavior of $\sigma(V_g)$ at high carrier density. For $V_g$ not too near $V_{g,min}$, the conductivity can be fit (see Figure 2) by

$$\sigma(V_g) = \begin{cases} \mu_e c_g (V_g - V_{g,min}) + \sigma_{res} & V_g > V_{g,min} \\ -\mu_h c_g (V_g - V_{g,min}) + \sigma_{res} & V_g < V_{g,min} \end{cases} \quad (1)$$

where $\mu_e$ and $\mu_h$ are the electron and hole field-effect mobilities, and $c_g$ is the gate capacitance per unit area, $1.15 \times 10^{-4}$ F/m$^2$, and $\sigma_{res}$ is the residual conductivity which is determined by the fit. The mobilities are reduced by an order of magnitude during each run, and recover upon annealing. The electron mobilities ranged from 0.081 to 1.32 m$^2$/Vs over the four runs, spanning over an order of magnitude, and nearly covering the range of mobilities reported to date in the literature (~0.1 to 2 m$^2$/Vs)[15].

The mobility is expected to depend inversely on the density of charged impurities $1/\mu \propto n_{imp}$. We assume $n_{imp}$ varies linearly with dosing time $t$ as potassium is added to the device. In Figure 3 we plot $1/\mu_e$ and $1/\mu_h$ vs. $t$, which are linear, in agreement with $1/\mu \propto n_{imp}$. From this point we parameterize the data by $1/\mu_e$, identified as proportional to the impurity concentration (the data set for $\mu_e$ is more extensive than $\mu_h$ because of the limited $V_g$ range accessible experimentally). We use the theoretical prediction that the product of mobility and impurity concentration is a constant[2,3,8-10], $\mu n_{imp} = C$, where $C = 5 \times 10^{15}$ V$^{-1}$s$^{-1}$, with the linear fits in Figure 3 to obtain the dosing rate $dn_{imp}/dt = (2.6 \sim 3.2) \times 10^{15}$ m$^{-2}$s$^{-1}$, which is of the correct order of magnitude with respect to estimates from residual gas analysis of the K flux during evaporation (see Methods); a more precise verification of the magnitude of $C$ is given by the shift in $V_{g,min}$ (see below). The value corresponds to a maximum concentration of $(1.4 \sim 1.8) \times 10^{-3}$ potassium atoms per carbon atom for the largest dosing time (18s) used.

The inset to Figure 3 shows that, although the $\mu_e$ and $\mu_h$ are not identical, their ratio is fairly constant at $\mu_e/\mu_h = 0.83$. The smaller scattering for repulsive vs. attractive impurity-carrier interaction is a unique aspect of Dirac electrons. A recent calculation[11] gives $\mu_e/\mu_h = 0.37$ for an impurity charge $Z = 1$, however the asymmetry is expected to be reduced when screening by conduction electrons is included.

As K-dosing increases and mobility decreases, the linear behavior of $\sigma(V_g)$ (see Figure 2) associated with charged impurity scattering dominates, as predicted theoretically[9]. At the lowest K-dosing level, sub-linear behavior is observed for large $|V_g - V_{g,min}|$ as anticipated. The dependence of the conductivity on carrier density $n \propto |V_g - V_{g,min}|$ is expected to be $\sigma \propto n^a$ with $a = 1$ for charged impurities, and $a < 1$ for short-range and ripple scattering (see Supplementary Note). Adding conductivities in inverse according to Matthiessen's rule indicates that scattering other than by charged impurities will dominate at large $n$, with the crossover occurring at larger $n$ as $n_{imp}$ is increased[9]. A previous study[15] also found a similar trend toward more linear $\sigma(V_g)$ for devices with lower mobility. Thus, our data indicate that the variation in observed field effect mobilities of graphene devices[15] is determined by the level of unintentional charged impurities.

We now examine the shift of the curves in $V_g$. Figure 4 shows $V_{g,min}$ as a function of $1/\mu_e$. Run 1 differs from Runs 2-4, presumably due to irreversible changes as potassium reacts with charge traps on silicon oxide and/or edges and defects of the graphene sheet. After Run 1, subsequent runs are very repeatable, other than an increasing rigid shift to more negative voltage in the initial gate voltage of minimum conductivity. (The same distinction between the first and subsequent experiments is seen in Figure 5 as well.) One might expect that the minimum conductivity would occur at the induced carrier density which precisely neutralizes the charged impurity density $n = -Zn_{imp}$, or $\Delta V_{g,min} = -n_{imp}Ze/c_g$[24], where $e$ is the elementary charge, and $Ze$ is the charge of the potassium ion. This prediction is shown as the dashed line in Figure 4; the experimental data do not follow this trend. Adam, et al.[10] proposed that the minimum conductivity in fact occurs at the added carrier density $\bar{n}$ at which the average impurity potential is zero, i.e. $\Delta V_{g,min} = -\bar{n}e/c_g$, where $\bar{n}$ is a function of $n_{imp}$, the impurity spacing $d$ from the graphene plane, and the dielectric constant of the SiO$_2$ substrate. The theory also assumes that $Z = 1$; experimentally, a reasonable evaluation[25, 26] of $Z$ for potassium on graphite is ~0.7; the effect of reduced impurity charge has not been calculated. The lines in Fig. 4 are given by the *exact result* of Adam et al.[10], and follow an approximate power-law behavior of $\Delta V_{g,min} \propto n_{imp}{}^a$ with $a = 1.2$~1.3, which agrees well with experiment. The only adjustable parameter is the impurity-graphene distance $d$; we show the results for $d = 0.3$ nm (a reasonable value for the distance of potassium on graphene[26-28]), and $d = 1.0$ nm (the value used by Adam, et al. to fit the result found by other groups for as-prepared graphene on SiO$_2$). Since $\Delta V_{g,min}$ gives an independent estimate of $n_{imp}$, the quantitative agreement in Figure 4 verifies that $C = 5\times10^{15}$ V$^{-1}$s$^{-1}$, as expected theoretically. Importantly, the observed $C$ is not increased by increased screening (e.g. by adsorbed water), as reported previously[24].

We now turn to the behavior near the point of minimum conductivity. We define three quantities: $\sigma_{min}$ is the minimum observed conductivity; the residual conductivity $\sigma_{res}$ is the point at which the linear extrapolations of $\sigma(V_g)$ meet, given by Equation 1; and the width of the minimum region $\Delta V_g$ is the difference between the two values of $V_g$ for which $\sigma_{min} = \sigma(V_g)$ in Equation 1.

Figure 5a shows the conductivities $\sigma_{min}$ and $\sigma_{res}$ as a function of $1/\mu_e$, and Figure 5b shows the plateau width $\Delta V_g$ as a function of $1/\mu_e$. Also shown are the predictions from the theory of Adam et al.[10] for $\sigma_{min}$ and $\Delta V_g$. Finite $\sigma_{res}$ has been predicted theoretically[12, 13] for graphene with charged impurities; however, the magnitude has not been calculated. The minimum conductivity drops upon initial potassium dosing, and shows a broad minimum near $4e^2/h$ before gradually increasing with further exposure. Notably, the cleanest samples show $\sigma_{min}$ significantly greater than $4e^2/h$, and strongly dependent on charged impurity density, indicating that the universal behavior[14] of $\sigma_{min}$ associated with the Dirac point is not observed even in the cleanest samples. The irreversible change in the value of $\sigma_{min}$ between Run 1 and Runs 2-4 is larger than the entire variation within Runs 2-4. This difference between the initial and the subsequent runs indicates that the initial K-dosing and anneal cycle introduces other types of disorder (possibly short range scatterers induced by irreversible chemisorption of potassium on defects or reaction of potassium with adsorbates) that have a comparable or greater impact on the minimum conductivity than charged impurities. That, for some disorder conditions (Run 1), $\sigma_{min}$ varies significantly with $n_{imp}$, but for other conditions (Runs 2-4) $\sigma_{min}$ is nearly independent of $n_{imp}$ for a very broad range of doping, suggests that the substantial variations reported in the literature (*i.e.* some groups report that $\sigma_{min}$ is a universal value[14], while other groups observe variation in $\sigma_{min}$ from sample to sample[15]) are likely due to poor control of the chemical environment of the devices measured. The observed residual conductivity $\sigma_{res}$ is finite and surprisingly constant (see Figure 5a); it is only weakly dependent on doping, and shows little variation between the first run and subsequent runs. The change of plateau width with doping (see Figure 5b) agrees only qualitatively with the theory, which predicts somewhat larger values and a sublinear dependence on doping. However, the quantitative disagreements between experiment and theory in Figures 5a and 5b are connected: mobility, minimum conductivity, and residual conductivity determine plateau width.

In conclusion, the dependence of electronic transport properties of graphene on the density of charged impurities has been demonstrated by controlled potassium doping of clean graphene devices in UHV at low temperature. The minimum conductivity depends systematically on charged impurity density, decreasing upon initial doping, and reaching a minimum near $4e^2/h$ only for non-zero charged impurity density, indicating that the universal conductivity at the Dirac point[14, 19, 20] has not yet been probed experimentally. The high-carrier density conductivity is quantitatively consistent with theoretical predictions for charged impurity scattering in graphene[2, 3, 8-10, 12, 13]. The addition of charged impurities produces a more linear $\sigma(V_g)$, and reduces the mobility, with the constanct $C = \mu n_{imp} = 5\times 10^{15}$ $V^{-1}s^{-1}$, in excellent agreement with theory. The asymmetry for repulsive vs. attractive scattering predicted for massless Dirac quasiparticles[11] is observed for the first time. Finally, the minimum conductivity does not occur at the point at which the gate-induced carrier density neutralizes the impurity charge, but rather when the average impurity potential is zero. As such, our experiment indicates that the inhomogeneous charge carrier distribution produced by the impurity potential determines the minimum conductivity point[10].

Other observations indicate the need for fuller experimental and theoretical understanding. The irreversible changes in the behavior around $V_{g,min}$ between the first and subsequent doping runs indicate that the precise value of the minimum conductivity point depends on the interplay of more than one type of disorder, and hence cannot be explained by existing theories[2-5, 7, 9, 10, 12, 13]. An interesting new feature, the residual conductivity (the extrapolation of the linear gate-voltage-dependent conductivity to $V_{g,min}$), may point to physics beyond the simple Boltzmann transport picture[12, 13]. Further experiments including introducing short-range (neutral) scatterers to graphene will be useful in addressing these questions. Full understanding may require scanned-probe studies of graphene under well controlled environmental conditions[21], which can completely characterize the disorder due to defects, charged and neutral adsorbates, and ripples, as well as probe the electron scattering from each[29].

Acknowledgements: This work has been supported by the Laboratory for Physical Sciences (EDW), the U.S. Office of Naval Research grant no. N000140610882 (CJ, MSF), NSF grant no. CCF-06-34321 (MSF), and the NSF-UMD-MRSEC grant no. DMR 05-20471 (JHC). MI is supported by the Intelligence Community Postdoctoral Fellowship program. We are grateful to Shaffique Adam and Sankar Das Sarma for useful discussions, and Susan Beatty and Gary Rubloff for use of the micro-Raman spectrometer.

**Methods**

Graphene is obtained from Kish graphite by mechanical exfoliation[30] on 300nm $SiO_2$ over doped Si (back gate), with Au/Cr electrodes defined by electron-beam lithography (see Fig. 1a). Raman spectroscopy confirms that the samples are single layer graphene[22] (see Fig. 1b). After fabrication, the devices are annealed in $H_2$/Ar at 300 °C for 1 hour to remove resist residues[21]. Gas flows are 1700 ml/min ($H_2$) and 1900 ml/min (Ar) at 1 atm; gases are flowing throughout heating and cooling. The devices are mounted on a liquid helium cooled cold finger in an ultra-high vacuum (UHV) chamber, so that the temperature of the device can be controlled from 20K to 490K.

Following a vacuum bakeout, each device is annealed in UHV at 490K overnight to remove residual adsorbed gases. Experiments are carried out at pressures lower than $5 \times 10^{-10}$ torr and device temperature $T$ = 20 K. Potassium doping is accomplished by passing a current of 6.5A through a getter (SAES Getters Inc. http://www.saesgetters.com/) for 40 seconds before the shutter is opened for 2 seconds. The getter temperature during each potassium dosage was 763±5 K as measured by optical pyrometry. The stability of the potassium flux was monitored by a residual gas analyzer positioned off-axis and behind the sample. Correcting for the geometry factor, the RGA-reported K-pressure would correspond to a flux of approximately $5 \times 10^{14}$ $m^{-2}s^{-1}$ at the sample. Since the RGA has not been calibrated for potassium, the value cannot be used quantitatively, but does confirm the order of magnitude of the deposition rate. All measurements shown here were performed on one four-probe device shown in Figure 1a, though several two-probe devices showed similar behavior (see Supplementary Figures S1, S3-S5).

Conductivity $\sigma$ is determined from the measured four-probe sample resistance $R$ using $\sigma = (L/W)(1/R)$. Because the sample is not an ideal Hall bar, there is some uncertainty in the (constant) geometrical factor $L/W$. We estimate $L/W$ = 0.80 ± 0.09, where the error bars represent the 68% confidence level. This 11% uncertainty in $L/W$ translates into an 11% uncertainty in the vertical axes of Figures 2 and 3, the horizontal axes of Figures 4 and 5b, and both axes of Figure 5a. Such scale changes are comparable to the spread among different experimental runs, and do not alter the conclusions of the paper. Notably, the uncertainty represents a systematic error, so *relative* changes in e.g. the minimum conductivity with charged impurity density are still correct.

Best fits to Eqn. 1 were determined using a least square linear fit to the steepest regime in the $\sigma(V_g)$ curves. The steepest regime of the $\sigma(V_g)$ curves was determined by examining $d\sigma/dV_g$; the fit was performed over a 2 V interval in $V_g$ around the maximum of $d\sigma/dV_g$. Other criteria for determining the maximum field effect mobility give similar results. The random errors in both $\mu$ and $V_{g,min}$ from the fitting procedure were typically less than 4%. Errors bars are shown in Figure 5 for random errors in $\sigma_{res}$ and the plateau width.

**Figure Captions**

**Figure 1 Graphene Device.** **a**, Optical micrograph of the device. **b**, 633 nm micro-Raman shift spectrum acquired over the device area, with Lorentzian fit to the D' peak, confirming that the device is made from single-layer graphene (vertical scale is same throughout **b**).

**Figure 2 Potassium doping of graphene.** The conductivity ($\sigma$) vs. gate voltage ($V_g$) curves for the pristine sample and three different doping concentrations taken at 20K in ultra high vacuum are shown. Data are from Run 3. Lines are fits to Eqn. 1, and the crossing of the lines defines the points of the residual conductivity and the gate voltage at minimum conductivity ($\sigma_{res}$, $V_{g,min}$) for each data set.

**Figure 3 Inverse of electron mobility $1/\mu_e$ and hole mobility $1/\mu_h$ vs. doping time.** Lines are linear fits to all data points. Inset: The ratio of $\mu_e$ to $\mu_h$ vs. doping time. Data are from run 3 (same as Figure 2).

**Figure 4 Shift of minimum conductivity point with doping.** The gate voltage of minimum conductivity $V_{g,min}$ is shown as a function of inverse mobility, which is proportional to the impurity concentration. All four experimental runs are shown. Each data set has been shifted by a constant offset in $V_{g,min}$ in order to make $V_{g,min}(1/\mu_e \rightarrow 0) = 0$, to account for any rigid threshold shift. The offset (in volts) is -10, 3.1, 5.6, and 8.2 for the four runs, respectively, with the variation likely to be due to accumulation of K in the SiO$_2$ on successive experiments. The open dots are $V_{g,min}$ obtained directly from the $\sigma(V_g)$ curves rather than fits to Eqn. 1 because the linear regime of the hole side of these curves is not accessible due to heavy doping. The solid and short-dashed lines are from the theory of Adam et al.[10] for an impurity-graphene distance $d = 0.3$ nm (solid line) and $d = 1$ nm (short-dashed line), and approximately follow power laws with slopes 1.2 and 1.3, respectively. The long-dashed line shows the linear relationship $\Delta V_{g,min} = n_{imp}Ze/c_g$ where $n_{imp} = (5\times10^{15}\,\text{V}^{-1}\text{s}^{-1})/\mu$ and $Z = 1$.

**Figure 5 Change in behavior near minimum conductivity point with doping. a**, The minimum conductivity and the residual conductivity (defined in text) as a function of $1/\mu_e$. **b**, The plateau width $\Delta V_g$ as a function of $1/\mu_e$. In **a** and **b**, data from all four experimental runs are shown, as well as the theoretical predictions of the minimum conductivity and plateau width from Adam et al.[10] for $d = 0.3$ nm (solid line) and $d = 1$ nm (short-dashed line).

Figure 1

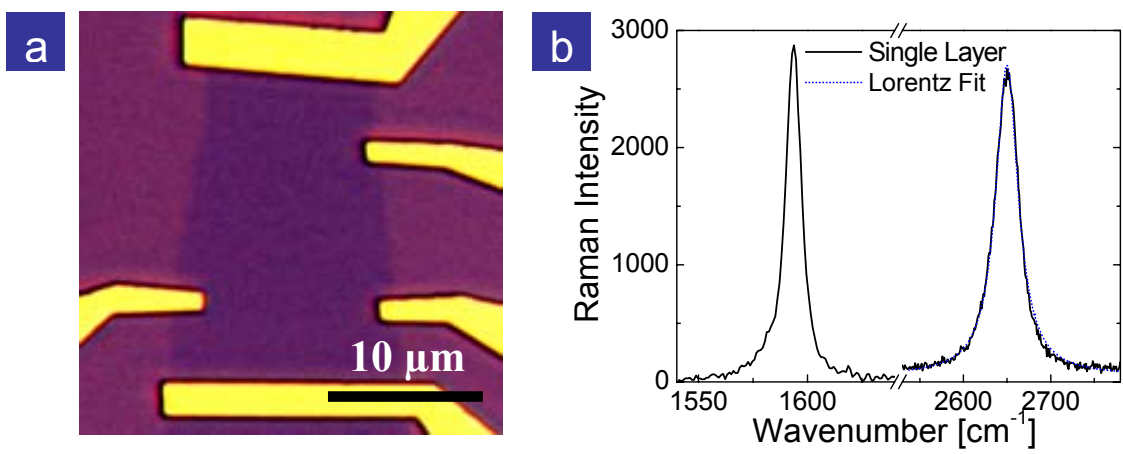

Figure 2

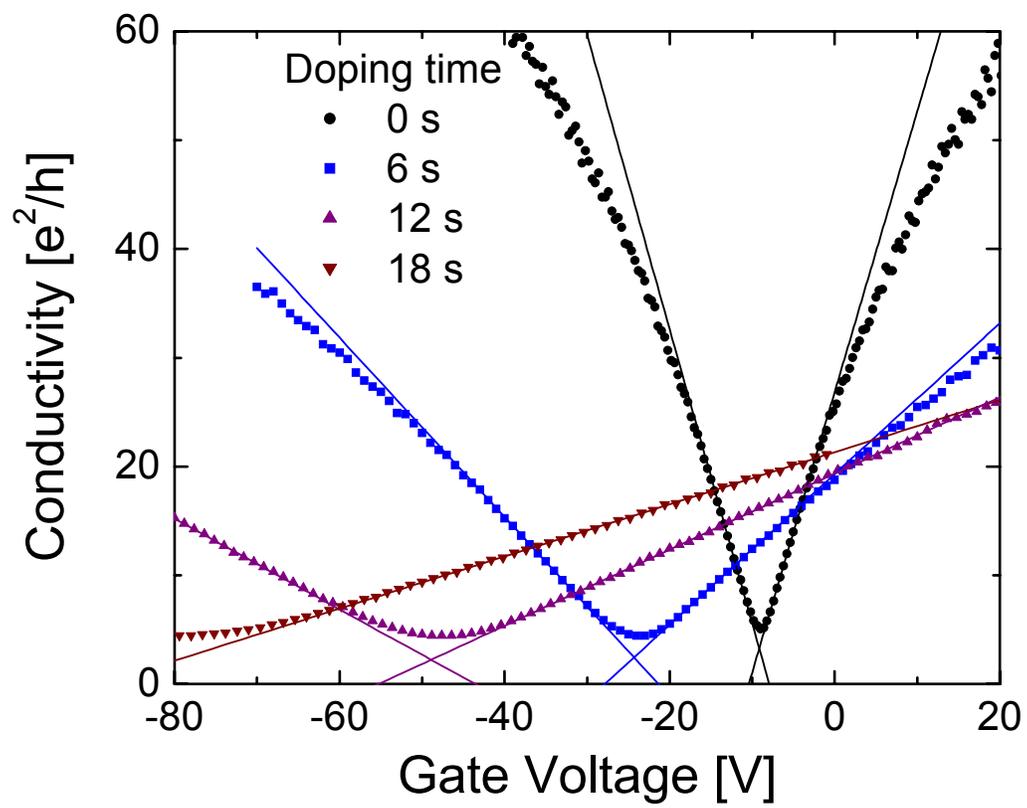

Figure 3

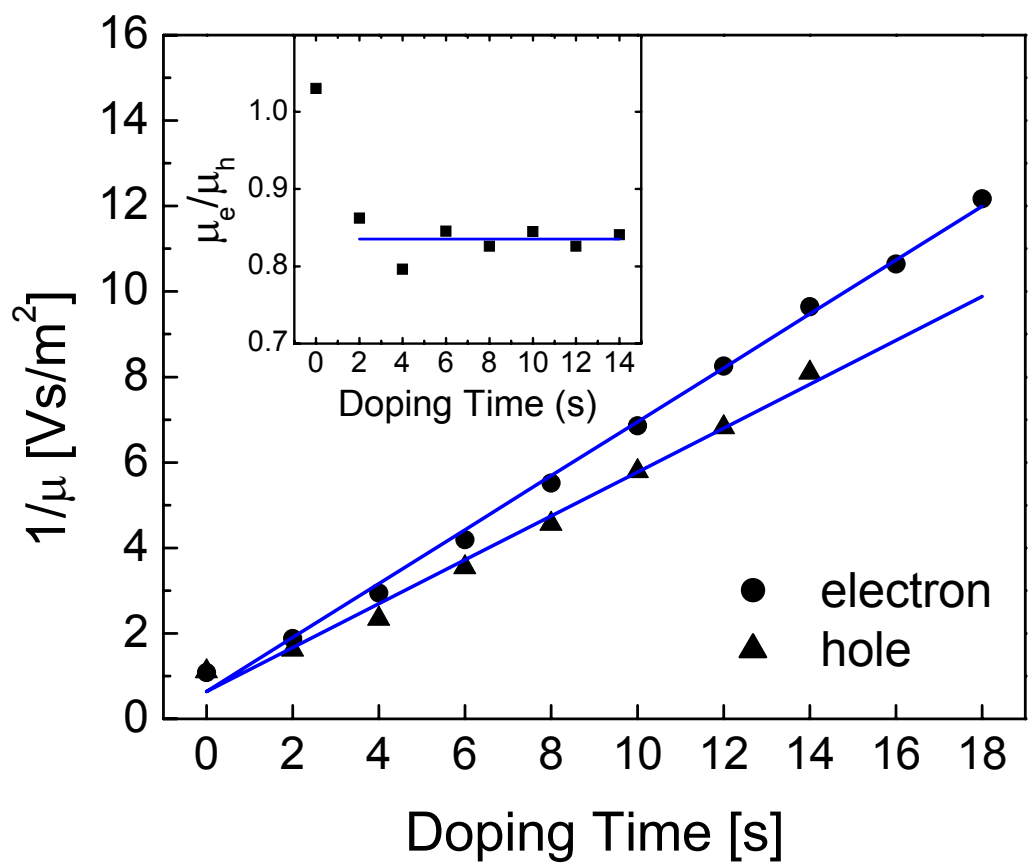

Figure 4

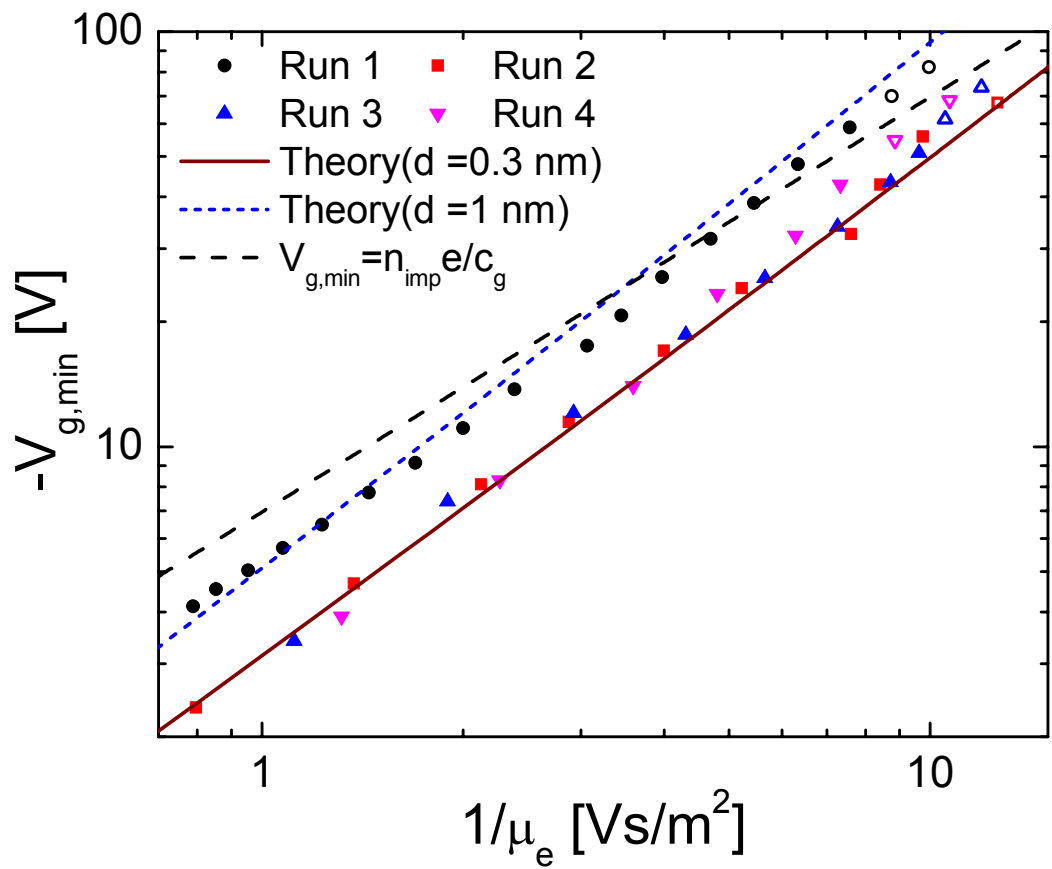

Figure 5

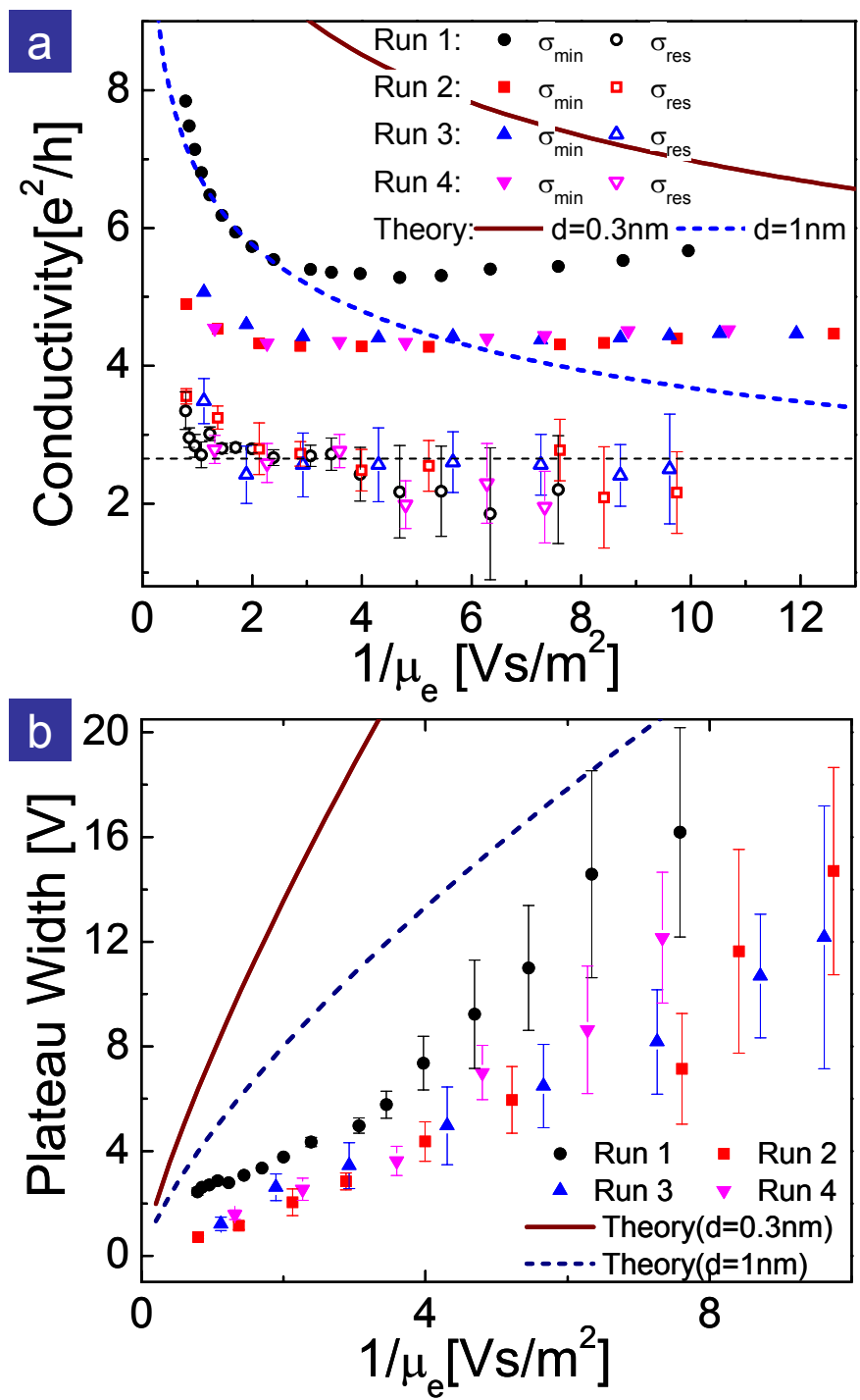